\documentclass[%
reprint,
superscriptaddress,
 amsmath,amssymb,
 aps,
]{revtex4-2}

\usepackage[utf8]{inputenc}
\usepackage{graphicx}% Include figure files
\usepackage{dcolumn}% Align table columns on decimal point
\usepackage{bm}% bold math
\usepackage{comment}
\usepackage{physics}
\usepackage[dvipsnames]{xcolor}

\newcommand{\rc}{R_\mathrm{C}}
\newcommand{\rl}{R_\mathrm{L}}
\newcommand{\acl}{\alpha_\mathrm{CL}}
\newcommand{\clmlp}{chlamylipo}
\newcommand{\clm}{\textit{Chlamydomonas}}

\setcitestyle{numbers,round,comma}



\begin{document}

\preprint{APS/123-QED}

\title{
Tuning microswimmer motility by liposome encapsulation: \\
a membrane clutch enables controlled swimming and cargo transport
%swimming and cargo transport of \textit{Chlamydomonas}-encapsulating liposome
}% Force line breaks with \\

\author{Koichiro Akiyama}
%\thanks{Equal contribution}
\affiliation{Graduate School of Science and Engineering, Hosei University, 3-7-2 Kajino, Koganei, 1848584, Japan}
\affiliation{These authors contributed equally to this work.}

\author{Sota Hamaguchi}%
%\thanks{Equal contribution}
\affiliation{Graduate School of Engineering Science, The University of Osaka, 1-3 Machikaneyama, Toyonaka, 5608531, Japan}
\affiliation{These authors contributed equally to this work.}

\author{Hiromasa Shiraiwa} 
\affiliation{Graduate School of Science and Engineering, Hosei University, 3-7-2 Kajino, Koganei, 1848584, Japan}

\author{Shunsuke Shiomi}
\affiliation{Graduate School of Science and Engineering, Hosei University, 3-7-2 Kajino, Koganei, 1848584, Japan}

\author{\\Tomoyuki Kaneko}
\affiliation{Graduate School of Science and Engineering, Hosei University, 3-7-2 Kajino, Koganei, 1848584, Japan}

\author{Masahito Hayashi}
%\email{masahito-hayashi@go.tuat.ac.jp}
\affiliation{Graduate School of Science and Engineering, Hosei University, 3-7-2 Kajino, Koganei, 1848584, Japan}
\affiliation{Department of Biotechnology and Life Science, Tokyo University of Agriculture and Technology, Koganei, Tokyo 1848588, Japan}
\affiliation{Corresponding author: masahito-hayashi@go.tuat.ac.jp}

\author{Daiki Matsunaga}
%\email{daiki.matsunaga.es@osaka-u.ac.jp}
\affiliation{Graduate School of Engineering Science, The University of Osaka, 1-3 Machikaneyama, Toyonaka, 5608531, Japan}

\affiliation{Corresponding author: daiki.matsunaga.es@osaka-u.ac.jp}

\date{\today}% It is always \today, today,

\begin{abstract}
Inspired by biology’s use of vesicles for targeted transport, many studies have propelled liposomes with active matter, creating synthetic systems that can be viewed as microscale biohybrid robots.
Nevertheless, the underlying motility mechanisms from a hydrodynamic perspective are often unresolved, and reliable velocity control remains challenging.
In the present study, we use our previously introduced biohybrid system termed chlamylipo, in which the motile alga \textit{Chlamydomonas reinhardtii} is encapsulated within a giant liposome, as a model system to investigate how liposome encapsulation regulates microswimmer motility under controlled perturbations.
From a fluid mechanical perspective, we establish a deformation--velocity expression that incorporates liposome radius, beating frequency, and membrane protrusion.
We further show that motility can be reversibly switched by incorporating light-responsive lipids, with the liposome acting as a ``clutch'' that modulates membrane protrusion.
To demonstrate its transport function, we further show cargo loading, carrying, and triggered release. These results show that liposome encapsulation can convert a motile microorganism into a controllable carrier, with the membrane acting as a clutch that regulates swimming and thereby cargo transport.
\end{abstract}

\begin{figure*}
    \centering
    \includegraphics[width=2.00\columnwidth]{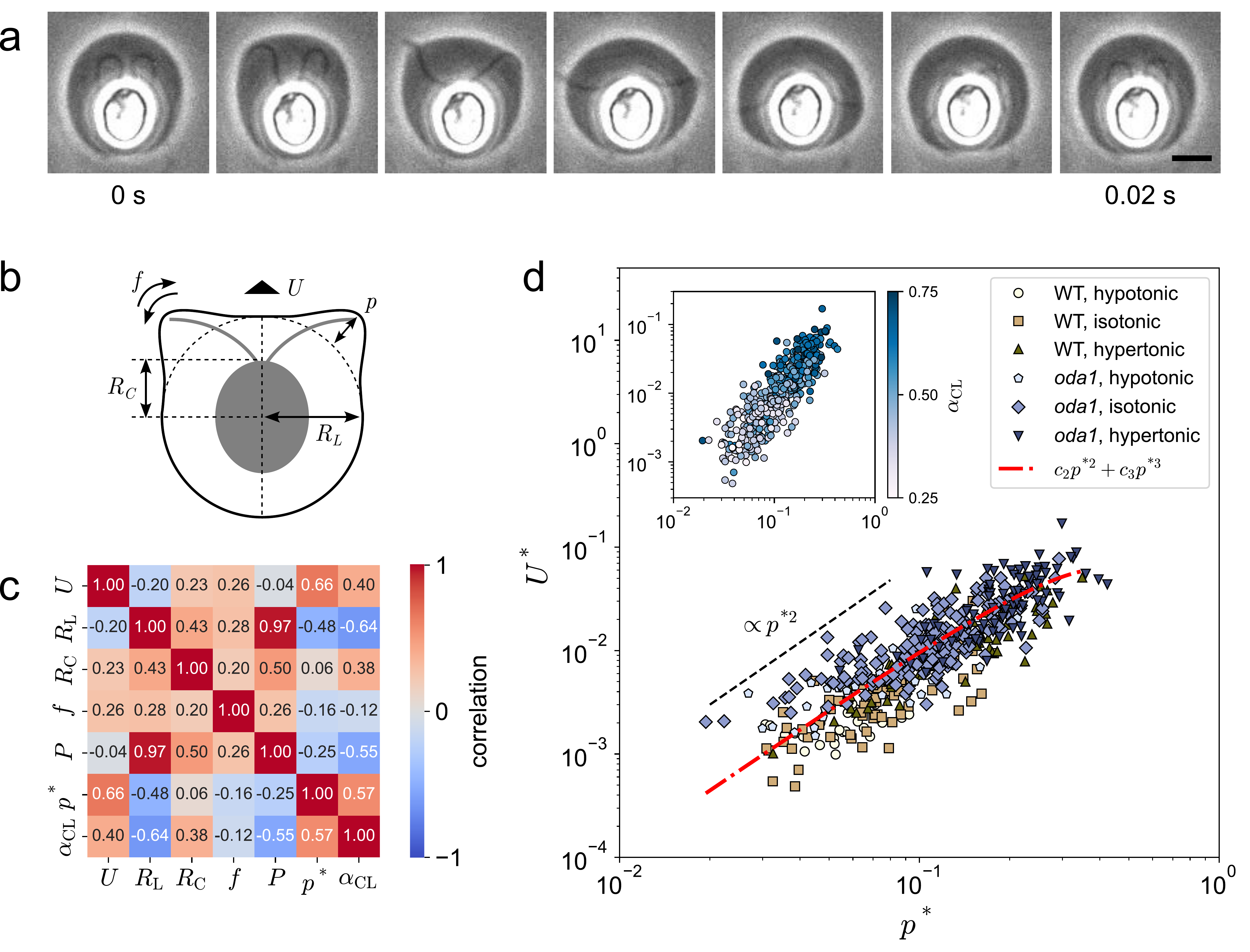}
    \caption{\textbf{Deformation--velocity relation of {\clmlp}.} \textbf{a} Time-lapse sequence of a single swimming stroke.  The black scale bar represents 5 $\mathrm{\mu m}$. \textbf{b} Schematic illustrating the variables extracted from image analysis. \textbf{c} Correlation matrix of the measured variables. \textbf{d} Deformation--velocity relation $p^*-U^*$ of {\clmlp}. The experimental conditions were varied by changing the {\clm} strains (wild-type CC-125 and \textit{oda1}), or by changing the osmotic pressure of the outer solution relative to the inner solution (hypotonic, isotonic, and hypertonic). The inset reports the same $p^*-U^*$ data, with points color-coded by \textit{Chlamydomonas}-to-liposome size ratio $\acl = \rc/\rl$. \label{fig:fig2}}
\end{figure*}

\maketitle

\section*{Introduction}
Mass transport at cellular and subcellular scales is fundamental to life, and living systems often use vesicles as mobile compartments for material delivery and intercellular signaling.
For example, cells transport intracellular vesicles over long distances along microtubules using motor proteins such as dynein and kinesin, internalize and secrete molecular cargo at the cell surface via endocytosis and exocytosis, and release extracellular vesicles that deliver proteins and RNAs to nearby and distant cells; red blood cells, in turn, ferry oxygen through the circulation.
To achieve artificial mass transport at the microscale, numerous studies in recent years have focused on the transport of liposomes, synthetic lipid vesicles, driven by active matter.
Previous studies have succeeded in transporting liposomes with the external actuation by bacteria \cite{kojima2012high,taherkhani2014covalent,nguyen2016active,felfoul2016magneto,dogra2016micro,alapan2018soft,akolpoglu2022magnetically}, algae \cite{wang2025light}, sperm \cite{geerts2014spermatozoa}, myosin \cite{lombardo2017myosin}, and kinesin \cite{bensel2024kinesin}.
These synthetic systems, which combine lipid membranes with living materials, can be regarded as microscale biohybrid robots; such robots integrate living organisms with engineered components to enable novel functionalities and precise control across scales \cite{webster2022biohybrid}.
Compared with nonliving synthetic systems driven by external fields \cite{ichikawa2001optical,pradhan2010targeted,talbot2017thermophoretic} or catalytic reactions \cite{somasundar2019positive,ghosh2019motility,borges2025minimal}, a potential advantage of biohybrid robots is that they can exploit biological functions intrinsic to the living component, including sustained motility and taxis.

In contrast to prior studies that relied on external active matter, self-propulsion has also been achieved by encapsulating active matter inside liposomes \cite{le2022encapsulated}.
In this previous work, motile \textit{Escherichia coli} (\textit{E. coli}) encapsulated in lipid vesicles deform the membrane into tubes that mechanically couple to flagellar motion, turning the tubes into propellers and enabling sustained vesicle swimming.
As a related direction beyond liposomes, droplet-based systems have likewise pursued propulsion by encapsulating active matter; theoretical frameworks has been proposed for the encapsulation of a microswimmer \cite{reigh2017swimming,aymen2023influence,della2023squirming,nganguia2025squirming} or many swimmers \cite{huang2020active} inside a droplet. 

Although these biohybrid robots show promise for biomedical applications \cite{carlsen2014bio,li2022biohybrid,zhang2024biohybrid,che2025engineering}, reliable motion control remains difficult because propulsion is driven by intrinsic biomotor activity with limited tunability.
Since motor activity is not readily adjusted by external cues, regulating swimming speed or reversibly switching motility on and off is a nontrivial task.
The relation between motor activity and motility from a fluid mechanical perspective is often lacking, which also makes reliable motion control difficult.
In our earlier short communication \cite{shiomi2025}, we proposed a biohybrid robot termed ``chlamylipo'', a giant liposome that encapsulates the motile alga \textit{Chlamydomonas reinhardtii}.
The {\clmlp} propels itself through time-irreversible deformation of the liposomal membrane, which is generated by the encapsulated {\clm} pushing the membrane with its flagella during breaststroke.
Because this biohybrid system retains phototaxis, the swimmer’s destination can be steered by the direction of external light.
Although the earlier study identified time irreversibility of the membrane deformation as an essential feature of propulsion, it remained unclear which geometric characteristics of the deformation govern the swimming velocity.
Establishing such a deformation--velocity relation could provide a fluid mechanical basis for controlling the motility of membrane-encapsulated biohybrid swimmers and, in turn, their cargo transport.
Therefore, the goal of this study is to use the {\clmlp} as a model system to establish a quantitative deformation--velocity relation and, based on this relation, to develop a means of regulating its motility through external stimuli.
We show that the liposomal membrane can act as a ``clutch'' that regulates swimmer motility, with this control extending to cargo transport.

\section*{Results}
\subsection*{Deformation--velocity relation of chlamylipo}

The chlamylipos were fabricated as described in our previous work \cite{shiomi2025}, by encapsulating {\clm} into giant unilamellar vesicles (GUVs) using the water-in-oil emulsion-transfer method \cite{pautot2003production,pautot2003engineering}.
This method has also been widely used to encapsulate other active materials, including actin filaments \cite{tanaka2018repetitive}, actomyosin networks \cite{litschel2021reconstitution,bashirzadeh2022encapsulated}, \textit{E. coli} \cite{morita2018direct,le2022encapsulated}, and bacterial cell division proteins \cite{litschel2018beating,ramirez2021ftsz}.
As reported in our previous work \cite{shiomi2025}, the {\clmlp} swims by cyclic, time-irreversible deformation of the liposome membrane driven by the {\clm} pushing the membrane with its flagella, as shown in Fig.~\ref{fig:fig2}a.
Note that this propulsion strategy, driven by cyclic surface shape deformation, is reminiscent of the envelope model for squirmers \cite{blake1971spherical}, and microswimmers in nature such as \textit{Volvox} \cite{pedley2016squirmers} and \textit{Eutreptiella/Euglena} \cite{arroyo2012reverse}.
The {\clmlp} exhibits intermittent forward swimming with a velocity of $U = 1.43 \pm 1.42$ $\mathrm{\mu m/s}$ (79 samples; wild-type strain CC-125 under isotonic outer solution), which is 1--2 orders of magnitude slower than free-swimming {\clm} cells \cite{minoura1995strikingly,leptos2009dynamics,fujita2014high}.

Although our earlier short communication \cite{shiomi2025} showed that the propulsion of {\clmlp} arises from time-irreversible cyclic deformation of the liposome membrane, it did not provide a quantitative framework linking the internally driven membrane deformation to the swimming velocity because of the limited number of observations $N=10$.
To understand the fluid mechanical basis of the deformation--velocity relation and establish a simple relation applicable over a broad parameter range, we perturbed the liposome deformation dynamics by varying the experimental conditions, allowing us to observe a wide variety of swimming modes.
The movements and deformations of {\clmlp} are observed under six distinct conditions by modifying the osmotic condition of the outer solution (hypotonic, isotonic, and hypertonic) and using two different {\clm} strains (wild-type CC-125 and mutant \textit{oda1}).
The osmotic condition was varied to alter the reduced volume of the liposome, as discussed in detail later.
The flagella-mutant strain \textit{oda1} is known for its slower beating of flagella, and its beating frequency and swimming velocity are nearly half of those of WT (CC-125) \cite{minoura1995strikingly, fujita2014high}. 
A schematic in Fig.~\ref{fig:fig2}b represents the obtained parameters via image processing; the swimming velocity $U = 2.10 \pm 2.55$ $\mathrm{\mu m}/s$, the beating frequency $f = 23.6 \pm 13.8$ Hz, the liposome radius $\rl = 7.73 \pm 1.75$ $\mathrm{\mu m}$, length of the major axis of {\clm} $\rc = 3.72 \pm 0.73$ $\mathrm{\mu m}$, and the protrusion $p = 0.90 \pm 0.45$ $\mathrm{\mu m}$ are obtained from the entire dataset $N=487$ with all six conditions.
The protrusion $p = P - \rl$ is evaluated by subtracting the liposome radius $\rl$ from the farthest distance $P$ from the centroid.
The dimensionless protrusion is defined as
\begin{equation}
  p^* = \frac{P - \rl}{\rl} = \frac{p}{\rl},
\end{equation}
and the value was $p^* = 0.13 \pm 0.07$; the result suggests that the protrusion $p$ is approximately $13\%$ of the liposome radius on average.
Now, can parameters describing membrane deformation be used to explain the observed swimming velocity?
Figure~\ref{fig:fig2}c shows the correlation matrix of the obtained variables, and the same figure with the scatter plots is shown in Fig.~S1.
While the velocity $U$ does not correlate with the farthest protrusion $P$, it strongly correlates with the dimensionless protrusion $p^*$ (correlation $r=0.66$), weakly with $f$ and $\rc$, and has weakly negative correlation with $\rl$.
Notably, as we will discuss later, there is a negative correlation between $p^*$ and $\rl$ and a weak negative correlation between $p^*$ and $f$.

\begin{figure*}
    \centering
    \includegraphics[width=2.00\columnwidth]{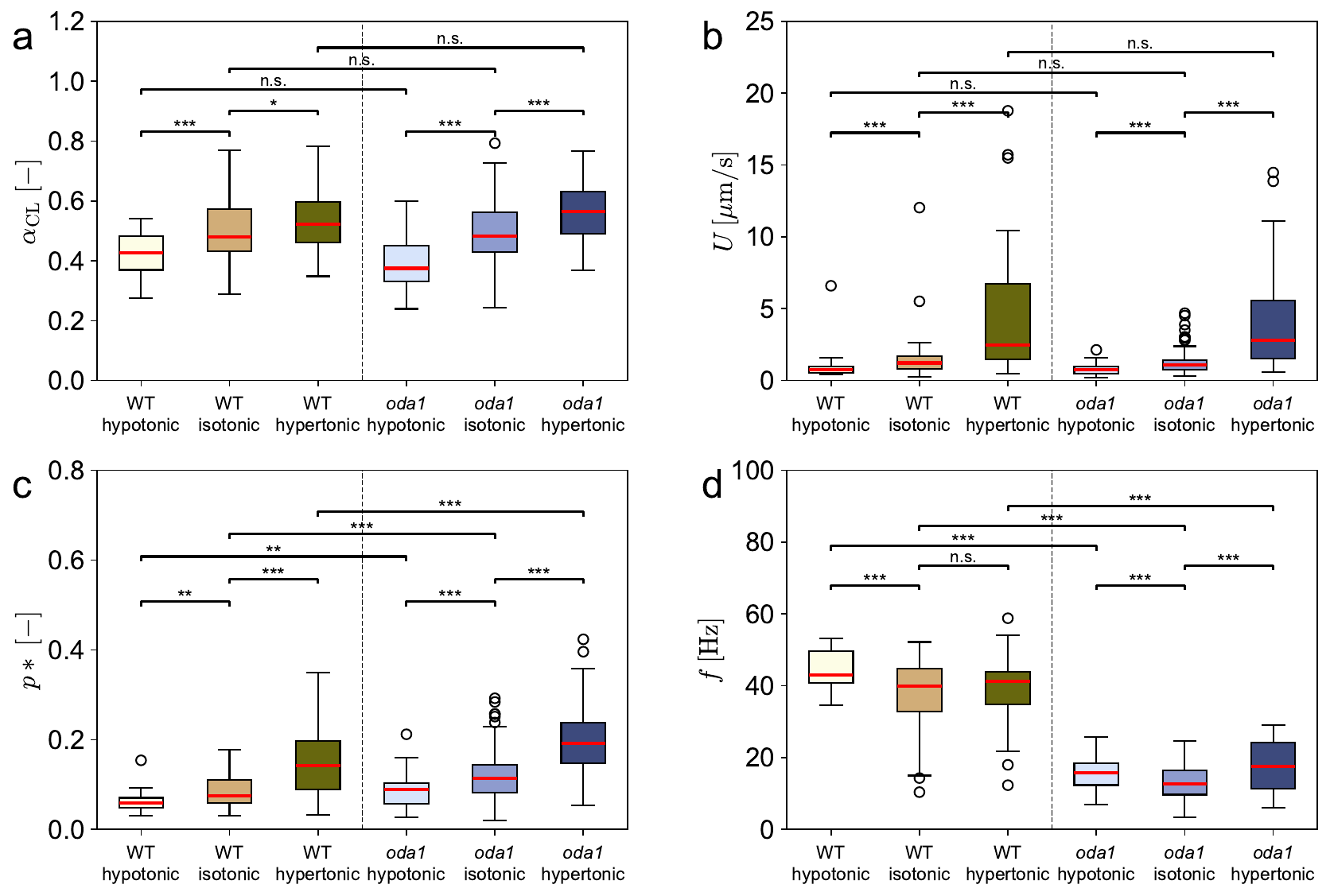}
    \caption{\textbf{Swimming characteristics under different experimental conditions.} Comparisons of \textbf{a} the \textit{Chlamydomonas}-to-liposome size ratio $\acl$, \textbf{b} the swimming velocity $U$, \textbf{c} the dimensionless protrusion $p^*$, and \textbf{d} the beating frequency $f$. Boxes denote the first and third quartiles, with the red horizontal line showing the median; whiskers extend to 1.5$\times$ the inter-quartile range, and any data beyond this range are plotted individually. Statistical significance was evaluated with a two-sided Mann-Whitney U-test (*** $p < 0.001$, ** $p < 0.01$, * $p < 0.05$). \label{fig:fig3}}
\end{figure*}

The correlation matrix shown in Fig.~\ref{fig:fig2}c does not readily provide a clear picture because the influence of each variable is entangled with that of the others. 
To compare data across different liposome sizes and beating frequencies, we now introduce dimensionless variables motivated by the classical small-amplitude envelope model \cite{lighthill1952squirming,blake1971spherical}.
The leading-order non-zero contribution to the time-averaged velocity scales with the square of the dimensionless deformation amplitude $p^*$ as $U \sim \rl f p^{*2} \sim f p^2/\rl$.
This form is also consistent with a simple scaling argument, in which $\rl f$ sets the characteristic velocity scale and the leading non-zero term for small periodic deformations is quadratic in $p^*$.
We therefore introduce a dimensionless velocity $U^* = U/(\rl f)$, giving the leading-order relation under small deformations as
\begin{equation}
  U^* = \frac{U}{\rl f} = c p^{*2} \label{eq:quadratic}
\end{equation}
where $c$ is a dimensionless prefactor.
Note that the dimensionless velocity $U^*$ can also be interpreted as the distance travelled per beat cycle normalized by the liposome radius.
Figure~\ref{fig:fig2}d shows that the $p^*-U^*$ data for the six different conditions collapse onto a single curve, and indeed $U^*$ follows the leading-order quadratic scaling $U^* \propto p^{*2}$ in the small deformations regime ($p^* \lesssim 0.1$).
Although theoretical analyses consistently establish a quadratic scaling in the small deformation regime, experimental observations of this scaling remain limited, except for a few examples, including those reported for natural \cite{pedley2016squirmers} and artificial swimmers \cite{leoni2009basic}.
The limited number of observations likely reflects the difficulty of varying the deformation level $p$, which is largely set by body geometry and stroke kinematics and is not easily tuned independently.
In this context, our measurements provide one of the first experimental demonstrations of the quadratic scaling with a sample size $n \sim O(10^2)$, and the present setup offers a practical model platform for probing the deformation--velocity relationship under controlled changes in $p$.

To examine how the leading-order quadratic scaling is modified at larger deformations, we considered the lowest-order correction motivated by a small-deformation expansion of the dimensionless velocity:
\begin{equation}
  U^* = \frac{U}{\rl f} = c_2 p^{*2} + c_3 p^{*3}.
\end{equation}
Here, the quadratic term represents the leading non-zero contribution \eqref{eq:quadratic}, whereas the cubic term provides the first correction beyond this leading-order regime, with the corresponding prefactors estimated from the entire dataset as $c_2 = 1.15$ and $c_3 = -1.91$.
The resulting curve, plotted in red in Fig.~\ref{fig:fig2}d, shows that the scaling slope becomes smaller than the quadratic once the deformation enters the large-deformation regime.
The negative prefactor of the cubic term $c_3 < 0$ indicates that this higher-order correction weakens the increase of $U^*$ with $p^*$ at large deformations, consistent with the theoretical prediction \cite{ishimoto2014swimming,wu2016amoeboid,ranganathan2018effect}.
In summary, the dimensionless analysis confirms that the classical leading-order quadratic scaling applies to the deformation--velocity relation of the {\clmlp}.
Neglecting the higher-order correction term $c_3 p^{*3}$, we use the relation
\begin{equation}
  U = c_2 \rl f p^{*2} = \frac{c_2 f p^2}{\rl} \label{eq:simple} 
\end{equation}
as a basis for the discussion in the following sections.

Up to this point, our discussion has primarily focused on building a deformation--velocity relationship using the ``outer" membrane properties ($\rl$, $f$, and $p$), but it is interesting to note here that the ``inner" configuration of {\clmlp} also affects the swimming velocity.
The inset of Fig.~\ref{fig:fig2}d shows the same $p^*-U^*$ relationship, but with points color-coded by the \textit{Chlamydomonas}-to-liposome size ratio $\acl = \rc/\rl$, and it shows that both the dimensionless protrusion $p^*$ and the dimensionless velocity $U^*$ are larger for larger $\acl$.
This relationship can be explained once we recognize that the deformation parameters are influenced by $\acl$; in other expressions $\rl(\acl)$, $f(\acl)$, $p(\acl)$, and consequently, $U(\acl)$. 
The protrusion $p^*$ has a positive correlation with $\acl$ (Fig.~\ref{fig:fig2}c; correlation $r=0.57$) since the tighter confinements leave less space between the cell and the membrane, allowing {\clm} to displace the membrane further outward.
The frequency $f$ has a weak negative correlation $r=-0.12$ with $\acl$ since tighter confinement distorts the flagellar stroke, elevates hydrodynamic friction, and thus slightly lowers the frequency.
The liposome size $\rl$ has a negative correlation $r=-0.64$ with $\acl$, which is obvious from the definition $\rl = \rc/\acl$.
Taken together, {\clmlp} with larger $\acl$ would protrude further and have smaller hydrodynamic friction due to its small radius $\rl$, thereby swims faster; as a result, the ratio $\acl$ has a strong positive correlation $r=0.40$ with the velocity $U$.
The results indicate that the membrane interior governs its deformation and, in turn, determines swimming velocity.

\subsection*{Velocity control via reduced volume}
Next, we take a closer look in Fig.~\ref{fig:fig3} how the swimming differs by the conditions, and show that the swimming characteristics of {\clmlp} can be tuned from outside.
We firstly discuss the effect of the reduced volume by changing the outer solution, and secondly, the effect of {\clm} strains.

The reduced volume $\nu$ ($0 < \nu < 1$) \cite{deuling1976curvature,helfrich1973elastic,seifert1997configurations} is a dimensionless parameter representing the ratio of the liposome's actual volume to that of a sphere with the same surface area, defined as $\nu = 6 \sqrt{\pi} V  A^{-3/2}$ where $V$ and $A$ are the volume and the surface area of liposomes, and this parameter shows the degrees of freedom available for membrane deformation.
When the reduced volume approaches unity, the liposome is maximally inflated and constrained to a nearly spherical shape.
Conversely, when this parameter is small, the excess surface area relative to the volume allows the liposome to deform into various shapes. 
Since the membrane deformation and protrusion are key factors for swimming, as revealed in the previous sections, we can now use the reduced volume as a controller to regulate motility.
In this work, the reduced volume of {\clmlp} is altered by changing the osmotic condition of the outer solution \cite{boroske1981osmotic}.
Figure~\ref{fig:fig3}a compares the \textit{Chlamydomonas}-to-liposome size ratio $\acl$ under different conditions. 
By altering the outer medium to a hypertonic/hypotonic solution, the liposome radius $\rl$ decreases/increases since the osmotic pressure difference drives outward/inward water flux across the membrane.
Therefore, as shown in the figure, the resultant $\acl$ becomes larger in a hypertonic outer solution and smaller in a hypotonic one; in other words, $\nu$ is smaller in a hypertonic outer solution and larger in a hypotonic one.
Due to this modification, the swimming velocity changes drastically as shown in Fig.~\ref{fig:fig3}b; under a hypertonic outer solution, the velocity is 3.32 times that of the isotonic case, whereas it slows to 0.69 of the isotonic case in a hypotonic solution.
This velocity difference arises since the dimensionless protrusion $p^*$ is drastically different, as shown in Fig.~\ref{fig:fig3}c.
Under the hypertonic outer solution, {\clm} can easily deform the membrane into various shapes owing to the small reduced volume, and this enhanced deformability leads to larger $p^*$ and $U$.
On the other hand, under the hypotonic solution {\clm} can only partially deform the nearly spherical membrane, resulting in smaller $p^*$ and $U$. 
Note that the beating frequency $f$ has a decreasing trend with the increase of $\acl$, as discussed in the previous section. 
To summarize, the swimming velocity can be modulated by varying the effective degrees of freedom for membrane deformation. 

Secondly, we examine how the velocity differs by {\clm} strains.
As mentioned earlier, previous studies reported that both the beating frequency and the resultant swimming velocity of the mutant \textit{oda1} are nearly half of those of WT \cite{minoura1995strikingly,fujita2014high}.
In agreement with these previous studies, the average beating frequency of WT (based on the data with all three osmotic conditions) was $f_\mathrm{WT} = 39.6 \pm 9.2$ Hz while that of \textit{oda1} was $f_\mathrm{oda1} = 14.8 \pm 5.6$ Hz, as shown in Fig.~\ref{fig:fig3}d.
Note that this beating frequency is slow compared to the bulk swimming, as the frequency is $f_\mathrm{WT} = 50-70$ Hz and $f_\mathrm{oda1} = 30-40$ Hz under bulk \cite{minoura1995strikingly,fujita2014high}, and it can be hypothesized that the viscous friction of flagella during the effective stroke increased by scratching the liposome membrane.
Although it was expected that the swimming velocity $U$ would also be slow for {\clmlp} with \textit{oda1}, no significance with WT-{\clmlp} in the velocity was found regardless of the osmotic condition, as shown in Fig.~\ref{fig:fig3}b; this result suggests that the swimmer which is slow under the bulk solution, somehow becomes comparable with the faster swimmer once embedded as the {\clmlp}.
The result appears counterintuitive since the swimming velocity $U$ should be proportional to the beating frequency $f$, as also shown in Eq.~\eqref{eq:simple}. 
This comparable velocity can be explained by the large protrusion $p^*$ of \textit{oda1}, shown in Fig.~\ref{fig:fig3}c, compensated for the velocity drop originating from $f_\mathrm{oda1} < f_\mathrm{WT}$.
Two hypotheses can be proposed to account for the enlarged protrusion observed in \textit{oda1}.
The first hypothesis is that the membrane’s relaxation time is so long that the flagellar beating is too rapid to displace the membrane sufficiently.
To elucidate the present membrane-flagella interaction, we introduce the following simple spring-damper model.
The time evolution of the protrusion $p$ can be modeled with an equation
\begin{equation}
  \eta \dot{p} + kp = F_0 \sin (2 \pi f t)
\end{equation}
where $k$ is the spring constant of the protrusion, $\eta$ is the hydrodynamic drag coefficient of the membrane recovery, $F_0$ is the amplitude of the flagella force, and $f$ is the beating frequency. 
By solving the steady-state solution, the amplitude of the protrusion is obtained as
\begin{equation}
  p_0 = \frac{F_0}{k \sqrt{1 + (2 \pi f \eta/k)^2}} = \frac{F_0}{k \sqrt{1 + (f\tau)^2}}, \label{eq:p0} 
\end{equation}
where $\tau = 2 \pi \eta/k$ is the relaxation time of the protrusion.
The equation shows that the protrusion becomes smaller for faster beating frequencies, which is in agreement with the experimental results.
Note that a simple toy model for the swimming velocity can be obtained from Eqs. \eqref{eq:simple} and \eqref{eq:p0} as follows:
\begin{equation}
  U = \frac{c_2 F_0^2}{\rl k^2} \cdot \frac{f}{1 + (f \tau)^2}
\end{equation}
and this expression suggests the existence of an optimal beating frequency $f_\mathrm{opt} = 1/\tau$ that maximizes the velocity $U$.
The second hypothesis is based on the differences in the beating pattern.
Previous work \cite{geyer2022ciliary} reports that the \textit{oda1} flagellar waveform has lower parabolicity than WT, which may contribute to the large protrusion observed in \textit{oda1}. 
A more detailed analysis of flagellum-membrane interactions would be an interesting direction for future work.

\begin{figure*}
    \centering
    \includegraphics[width=2.0\columnwidth]{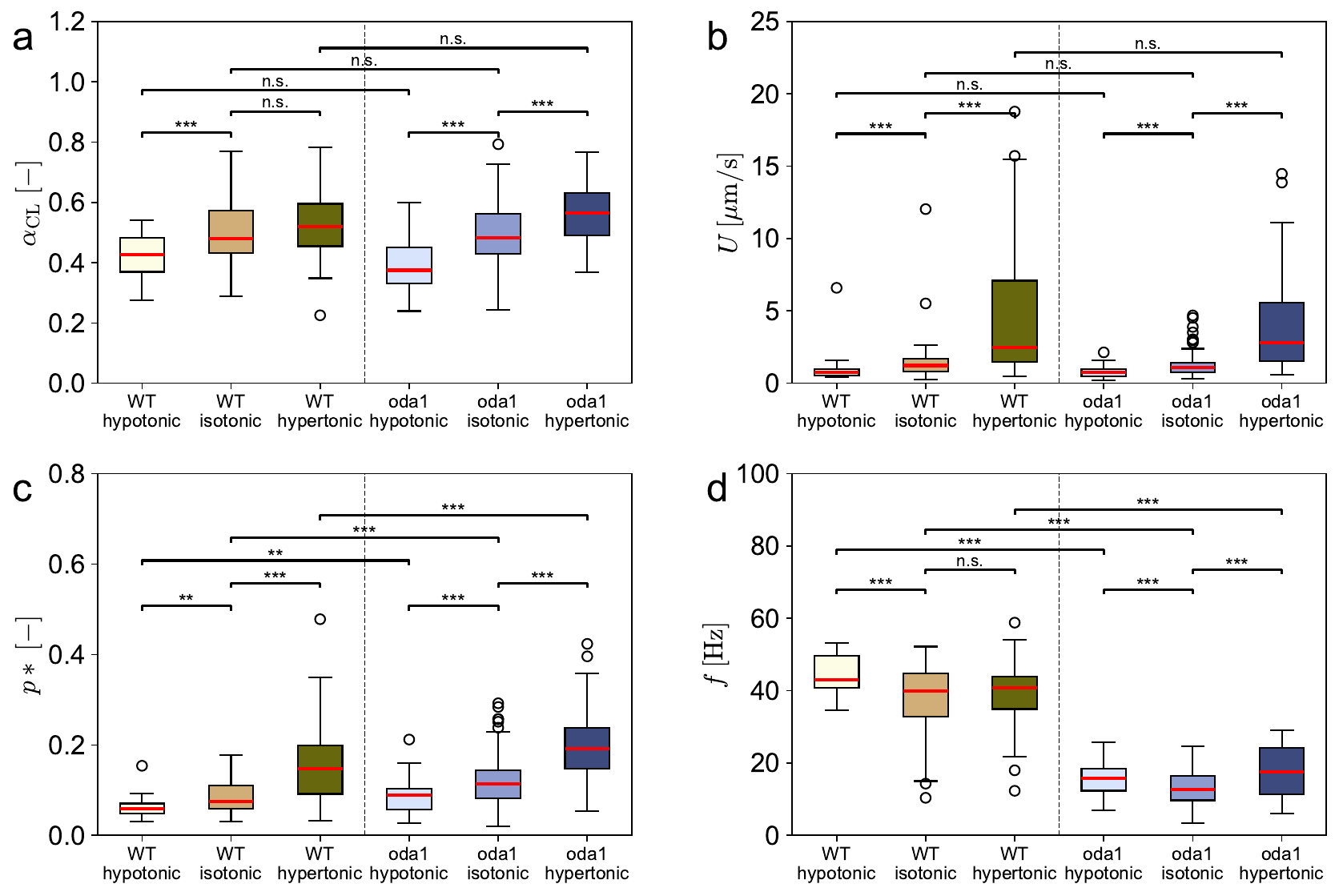}
    \caption{\textbf{Photoswitchable motility by utilizing membrane as a ``clutch''.}
    \textbf{a} Schematic figure of the reversible motility photoswitching. The {\clmlp} becomes motile with a UV pulse, while it can not swim after the blue light pulse.
    \textbf{b} Time history of the distance from the initial position $|\mathbf{x}(t) - \mathbf{x}(0)|$ and the swimming velocity $U$ from Supplemental Movie S2. Blue and purple vertical lines show the time of the blue light and UV pulses, respectively.
    \textbf{c} Time-lapsed snapshots from Supplemental Movie S2. The black scale bar represents 20 $\mathrm{\mu m}$.
    \textbf{d} Drawing letters ``CL" by manipulating the {\clmlp} trajectory with lights. The trajectory of the top figure is colored with the elapsed time, while the bottom figure is colored with the swimming velocity. The scatter-marker radius is inversely proportional to the velocity magnitude. The black scale bar represents 50 $\mathrm{\mu m}$.
    \textbf{e}--\textbf{f} Photoswitching of membrane protrusion in liposomes encapsulating \textit{E. coli} and magnetic beads, respectively. The NdFeB magnet was positioned to the right of the field of view in \textbf{f}. The magnetic field strength at the sample position was approximately 3 mT. 
    \label{fig:fig4}}
\end{figure*}

\begin{figure*}
    \centering
    \includegraphics[width=1.8\columnwidth]{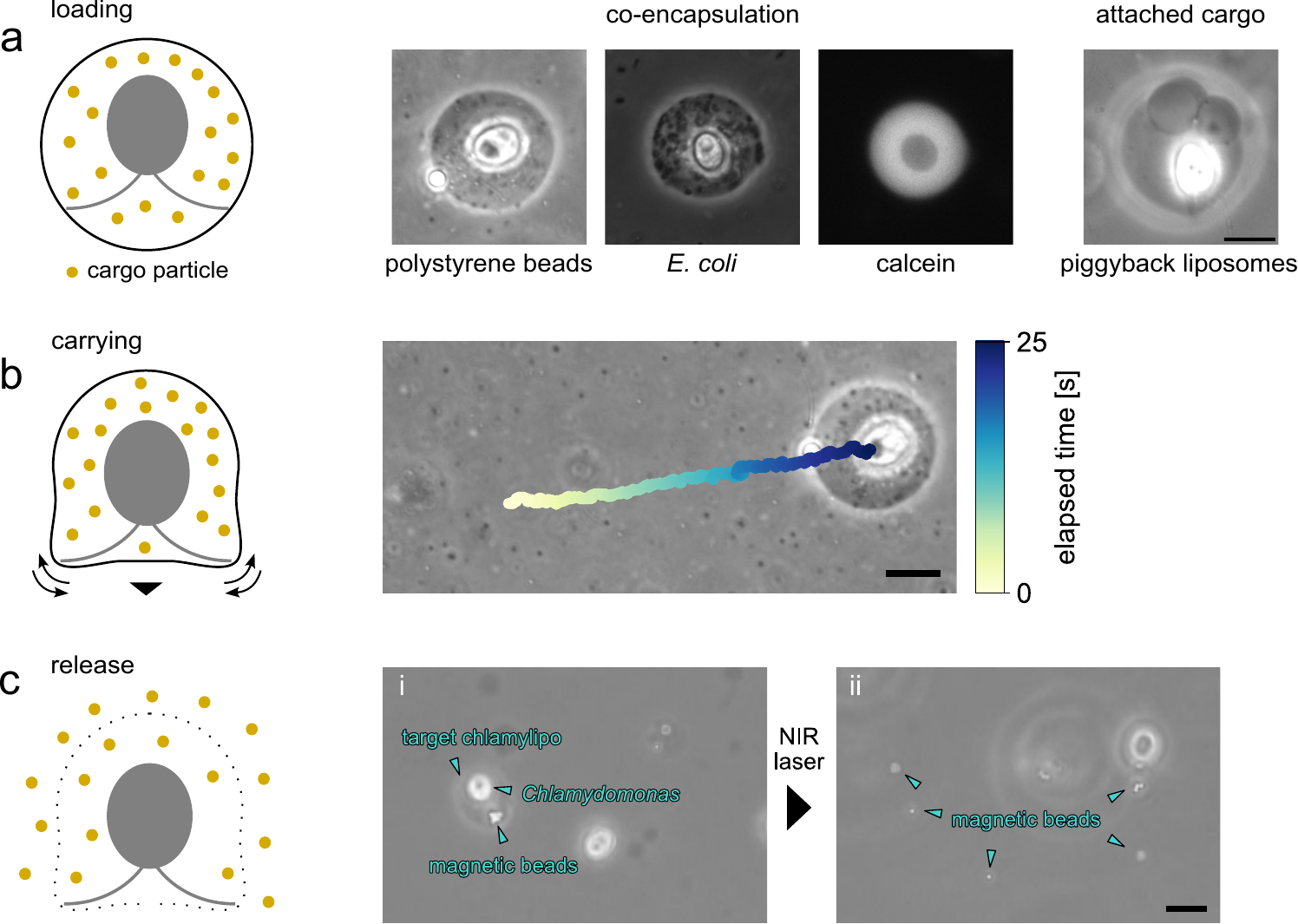}
    \caption{\textbf{Cargo loading, carrying and releasing of {\clmlp}.}
    \textbf{a} Representative snapshots of cargo-loaded {\clmlp}s. The cargos were co-encapsulated with {\clm}, except in the rightmost image, where small liposomes were attached to the exterior of the {\clmlp}.
    \textbf{b} Snapshots of swimming {\clmlp} containing micro-sized beads co-encapsulated with {\clm} as cargo. Images are taken from Supplemental Movie S8.
    \textbf{c} Cargo release from {\clmlp} upon NIR laser irradiation. Co-encapsulated magnetic beads (i) were expelled as the liposome membrane burst, followed by their diffusion away (ii). Images are taken from Supplemental Movie S9.
    All black scale bars represent 10 $\mathrm{\mu m}$. \label{fig:fig5}}
\end{figure*}

\subsection*{Photoswitchable motility with membrane ``clutch"}
From the preceding analysis, we find that the protrusion $p$ plays an essential role in determining the swimming velocity $U$, and that the protrusion amplitude can be controlled via the reduced volume $\nu$.
Although the osmotic condition can modify the reduced volume, this approach is not a practical method to achieve cyclic regulation of the swimming velocity.
To achieve instantaneous control of motility, we introduce the photoswitchable phospholipid AzoPC (azobenzene-containing phosphatidylcholine) into the liposomal membrane \cite{pernpeintner2017light}, enabling light-based control of the reduced volume.
AzoPC is a photoswitchable phospholipid that undergoes light-induced photoisomerization: blue-light irradiation stabilizes the rod-like \textit{trans} state, whereas UV irradiation drives the bent/cone-like \textit{cis} state, as schematically illustrated in Fig.~\ref{fig:fig4}a.
As a consequence of this conformational change at the lipid level, the liposome surface area increases upon UV irradiation and decreases upon blue-light irradiation, thereby allowing the reduced volume to be modulated by external light \cite{pernpeintner2017light,pritzl2022postsynthetic,aleksanyan2023photomanipulation,manafirad2023structural}.
In our {\clmlp} system, a hypotonic outer solution was used to osmotically inflate the liposome before UV irradiation, such that its initial reduced volume was expected to be close to unity.
For our membrane composition with DOPC:AzoPC = 1:1, the membrane area is expected to increase by approximately 5\% upon UV irradiation \cite{manafirad2023structural}, which corresponds to a decrease in reduced volume from $\nu \simeq 1$ to $\nu \simeq 0.93$ if the enclosed volume is assumed to remain unchanged.
We note here that, unlike in the osmotic experiments, $\acl$ is not a suitable descriptor for tracking the expected change in reduced volume $\nu$, since $\acl=\rc/\rl$ does not directly capture the photoinduced change in membrane area.
Consistent with this expected decrease in reduced volume, the inner {\clm} can protrude the membrane outward after a UV pulse, whereas the liposome remains nearly spherical after a blue-light pulse, as shown in the bottom images of Fig.~\ref{fig:fig4}a and Supplemental Movie S1.
It is worth noting that the {\clm} continues to beat its flagella regardless of the membrane state; the membrane therefore functions as a ``clutch" that regulates how much of this internal motion is transmitted to the exterior.
This reversible switching of the clutch enables us to control the {\clmlp} motility on and off.
Note it can be also said that changing the motility level effectively modifies the P\'{e}clet number, which quantifies the relative importance of advective swimming to the liposome’s diffusive motion.

Figures~\ref{fig:fig4}b--c and Supplemental Movie S2 show the {\clmlp} movement under alternating UV and blue light pulses at every 20~s intervals; the irradiating time of each pulse was approximately 500~ms. 
Note that we used the green light to guide the {\clmlp} towards the right side, UV and blue light to switch motility, and red light is used for observation, as {\clm} exhibits no phototactic response to red light.
The top panel of Fig.~\ref{fig:fig4}b shows the distance from the initial position $|\mathbf{x}(t) - \mathbf{x}(0)|$ where $\mathbf{x}(t)$ is the position at time $t$, while the bottom panel shows the swimming velocity $U$.
Although the {\clmlp} is in the non-motile state at the initial condition (i), it begins to swim forward quickly after the first UV pulse (ii) with a velocity $\approx 2.0$~$\mathrm{\mu m/s}$, and returns to the non-motile state right after the first blue-light pulse (iii).
These motile and non-motile states can be switched reversibly and cyclically, and we have successfully repeated this clutch switching more than seven times, as shown in the figures.
These two states enable precise manipulation of the swimmer’s motion; Fig.~\ref{fig:fig4}d and Supplemental Movie S3 show that we can trace letters ``CL", the initials of {\clmlp}, by steering the position.
The top image of Fig~\ref{fig:fig4}d shows the trajectory colored with the elapsed time, whereas the bottom image is colored with the swimming velocity, with marker radius inversely proportional to the local velocity magnitude.
During tracing, after the swimmer reached each corner of CL letters, the {\clmlp} was switched to the non-motile state by blue light; these pauses are indicated by large blue markers.
The green light position was then shifted to set the next heading before switching back to the motile state.
While steering is possible without the clutch, access to motile and non-motile states allowed precise cornering; pausing at each vertex of the ``CL" trajectory enabled accurate reorientation and precise letter formation.
Although this tracing is primarily a demonstration, the ability to pause the swimmer in place could be useful in several practical applications, as discussed in the Discussion.
We also note that motility can be switched collectively by broadcasting the switching signal to all {\clmlp}s, or individually by selectively irradiating only the target {\clmlp}, as shown in Fig.~S2 and Supplemental Movie S4.

Importantly, the applicability of this photoswitching strategy is not limited to {\clm}, but extends to liposomes containing other active or externally driven agents.
For example, the tube-like membrane protrusion generated by the swimming of encapsulated \textit{E. coli} \cite{le2022encapsulated} can be reversibly switched on and off by light, as shown in Fig.~\ref{fig:fig4}e and Supplemental Movie S5.
Similarly, encapsulated magnetic beads form a protrusion toward an externally applied magnet, and the protrusion length can be reversibly modulated by photoswitching of the membrane, as shown in Fig.~\ref{fig:fig4}f and Supplemental Movie S6.
Note that this switching could also be achieved by varying the magnetic field gradient; this experiment is intended to demonstrate photoswitchable control of membrane protrusion rather than an advantage over direct magnetic control.
These results suggest that the liposome membrane can serve as a tunable mechanical transmitter, regulating how strongly the activity or forcing generated by an encapsulated agent is expressed as membrane deformation.
As discussed later, this capability may be useful in active matter research since it provides a means of modulating the effective activity of the agent.

\subsection*{Cargo transport with {\clmlp}}
Finally, we examined whether the {\clmlp} could function as a self-propelled liposomal transport system through cargo loading, carrying, and triggered release.
The {\clmlp} that co-encapsulates cargo can be fabricated by preparing an aqueous solution containing both {\clm} and the target cargo during water-in-oil emulsion formation.
Various types of cargo can be co-encapsulated as shown in Fig.~\ref{fig:fig5}a and Supplemental Movie S7, including particulate cargo such as beads and \textit{E. coli}, as well as soluble molecular cargo such as the fluorescent dye calcein.
Although the configuration differs from co-encapsulation, small liposomes attached to the exterior of the {\clmlp} can also be regarded as externally loaded cargo.
The {\clmlp} can swim while carrying cargo as shown in the movie, and the swimming velocities measured for all tested cargo types remained on the order of $\mathrm{\mu m/s}$, as in the cargo-free case.
The {\clmlp} also retained phototaxis and exhibited persistent swimming under cargo co-encapsulation, as demonstrated in Fig.~\ref{fig:fig5}b and Supplemental Movie S8, which show a {\clmlp} swimming while carrying 0.5~$\mathrm{\mu m}$ polystyrene beads.
Although these results demonstrate that motility is maintained while carrying cargo, it should be noted that the choice of $\acl$ when designing the {\clmlp} involves a geometric trade-off: the {\clmlp} swims faster at larger $\acl$, whereas a larger internal volume is available for cargo at smaller $\acl$.
After guiding the {\clmlp} to the destination, the inner cargo can be released from the membrane using an NIR (near-infrared) laser \cite{shiraiwa2026}, as shown in Fig.~\ref{fig:fig5}c and Supplemental Movie S9.
The snapshot images show a {\clmlp} containing 1.0~$\mathrm{\mu m}$ magnetic beads (i) that was irradiated with an NIR laser, leading to the release of the beads, which subsequently diffused away (ii).
The magnetic beads were expelled because NIR-induced local heating increased the membrane fluidity, leading to transient pore formation and eventual membrane bursting \cite{paasonen2007gold,wu2008remotely,lajunen2015light}.
Although the choice of cargo is constrained by its compatibility with {\clm}, these results, together with the membrane-mediated motility control demonstrated above, support the concept of using membrane encapsulation to create controllable liposomal carrier capable of cargo loading, carrying to a target location, and triggered release.

\section*{Discussion}
In this study, we analysed chlamylipo swimming from a fluid mechanical perspective and showed how the liposomal membrane can act as a clutch to enable controlled swimming and cargo transport.
A nondimensional analysis indicated that the speed scales with the beating frequency, the square of the protrusion amplitude, and the inverse of the liposome radius.
The ratio of the {\clm} radius to the liposome radius also plays an important role: a larger ratio allows the membrane to protrude farther, leading to faster swimming.
Thus, if cargo capacity is not a priority and the goal is simply to obtain a faster {\clmlp}, the {\clm}-to-liposome size ratio should be set to a larger value.
During the analysis, we also found that the mutant \textit{oda1}, whose beating frequency and free-swimming velocity are about half those of the wild-type, reaches a speed comparable to the wild-type when encapsulated; the larger protrusion of \textit{oda1} appears to compensate for the lower frequency.
The swimming velocity can be modulated by tuning the protrusion level in response to external stimuli, such as osmotic conditions and light switching. 
When the inner volume of the liposome increases due to the osmotic difference across the membrane, the {\clmlp} swims poorly as the membrane has only limited degrees of freedom for deformation.
When the inner volume decreases, on the other hand, the {\clmlp} can swim faster thanks to the rich deformability of the membrane.
By introducing the photoswitchable phospholipid AzoPC into the liposomal membrane, the motility can be reversibly controlled with light, and the membrane works as a ``clutch".
The {\clmlp} switches to a motile state after UV illumination as the effective membrane area increases, and it returns to a non-motile state after subsequent blue light exposure.
Although the light-based control used here was a simple two-state switch, finer continuous tuning should be achievable in future work by more finely controlling the membrane area.
We also demonstrated that the {\clmlp} can function as a liposomal carrier by transporting encapsulated cargo and enabling its triggered release.
The present UV/blue-light control is primarily suited to micromanipulation in optically accessible environments, such as microfluidic devices, rather than to \textit{in vivo} applications requiring deeper light penetration.
Red light responsive phospholipids, red-azo-PC, have been developed \cite{pritzl2021optical}, and their use could both reduce phototoxicity by avoiding UV illumination and allow deeper light penetration.

As noted in the introduction, transport at the microscale remains a central challenge, and many studies have explored liposome- or droplet-based cargo delivery using either internally encapsulated active matter or external propulsion by surrounding active matter.
Compared with systems in which liposomes are attached to microswimmers \cite{kojima2012high,taherkhani2014covalent,nguyen2016active,felfoul2016magneto,dogra2016micro,alapan2018soft,akolpoglu2022magnetically,wang2025light}, encapsulating a microorganism inside a liposome entails a clear disadvantage: the membrane masks its native propulsion, resulting in a substantial reduction in swimming velocity relative to the free microorganism.
Encapsulation, however, also creates a distinctive functionality in the present system: rather than serving solely as a cargo-containing compartment, the liposome membrane acts as a clutch that regulates the transmission of biological activity to external motion, thereby enabling control of the swimming velocity.
In microorganism-driven biohybrid robots, previous studies have offered limited means to modulate velocity or to reversibly switch motility off, often relying instead on irreversible approaches such as laser-induced disabling of the microswimmer.
In this context, the reversible switching of motility demonstrated here adds two important functionalities to biohybrid robots. First, it enables cargo retention without transport, allowing the transport and release rates to be tuned. Second, it improves positioning precision by enabling the robot to be stopped and restarted near a desired location.
The internal actuator need not be limited to {\clm}, and the same approach could be extended to other biological or synthetic swimmers.
Biological actuators with intrinsic taxis, such as magnetotaxis or chemotaxis, could provide alternative means of navigation, and synthetic soft swimmers driven by self-sustained deformation may benefit from the membrane as an additional means of regulating their swimming speed.

Beyond its application to biohybrid robots, the idea of using a membrane as a clutch may also provide an experimental tool for active matter research. Active matter systems are often governed by the balance between activity and passive effects such as elasticity, viscosity, and confinement, yet activity itself is often difficult to tune experimentally. The ability to vary effective activity externally and reversibly may therefore help clarify how this balance shapes individual and collective dynamics. We expect this strategy to contribute both to the development of biohybrid robotics and to the fundamental understanding and control of active matter.

\section*{Materials and Methods} 
\subsection*{Reagents}
Phospholipids, including 1,2-dioleoyl-\textit{sn}-glycero-3-phosphocholine (DOPC) and azobenzene-containing phosphatidylcholine (AzoPC), were purchased from Avanti Polar Lipids (Alabaster, AL, USA).
Bovine serum albumin (BSA), glucose, sucrose, and calcein were purchased from FUJIFILM Wako Pure Chemical Corporation (Osaka, Japan).
Medium-chain triglyceride (MCT) oil (edible grade) was purchased from Nisshin OilliO Group (Tokyo, Japan).
Magnetic beads were purchased from Thermo Fisher Scientific (Waltham, MA, USA).
All other chemicals were of analytical grade and used as received.
Ultrapure water (Milli-Q, Millipore) was used to prepare all aqueous solutions.

\subsection*{Cell culture and preparation of inner solution and outer solution}
The wild-type \textit{Chlamydomonas reinhardtii} strain CC-125 and the mutants \textit{oda1} and \textit{fa1} were cultured photoautotrophically in TAPP medium with aeration at 25~$\mathrm{^\circ C}$ under a 12-h light/12-h dark cycle (approximately 100~$\mathrm{\mu mol}$ photons/($\mathrm{m^2,s}$)).
The \textit{fa1} strain was used in the experiments shown in Figs.~\ref{fig:fig4} and \ref{fig:fig5} because it displayed a swimming mode similar to that of CC-125 but appeared to exhibit a stronger phototactic response.
\textit{Chlamydomonas} cells in the logarithmic growth phase were harvested by brief centrifugation and resuspended in an inner solution (IS) composed of 30\% Percoll, 100 mM sucrose and 0.1 mM calcein.
Approximately $10^6$ cells were used for each encapsulation experiment.
The outer solution (OS) contained 100 mM glucose and 1 mg/mL bovine serum albumin (BSA).
Both IS and OS were kept at room temperature and used within one day of preparation.
Liposomes encapsulating magnetic or polystyrene beads were also prepared by suspending the beads in the IS.
A rectangular NdFeB permanent magnet (20 × 12 × 5 mm; MonotaRO, Japan) was used to manipulate the magnetic beads. The magnetic field strength at the sample position was approximately 3 mT.
\textit{Escherichia coli} (RP437) was cultured in LB medium.

\subsection*{Preparation of lipid solution}
Two types of lipid solutions were prepared for liposome formation: a standard lipid solution (LS) containing DOPC and a photoresponsive lipid solution (AzoLS) containing DOPC and AzoPC at a 1:1 molar ratio.
For each solution, the required lipids were dissolved in chloroform and transferred to a PCR tube.
The chloroform was removed by vacuum drying for 15~min, leaving a thin lipid film on the tube wall.
After drying, 100~$\mathrm{\mu L}$ of MCT oil was added to each tube.
LS contained 1~mM DOPC, whereas AzoLS contained 0.5~mM DOPC and 0.5~mM AzoPC, corresponding to a total lipid concentration of 1~mM.
Each tube was incubated at 70~$\mathrm{^\circ C}$ for 5~min using a thermal cycler and gently tapped to disperse the lipid film in the oil phase.
The heating and mixing procedure was repeated three times.
The resulting lipid solutions were stored at room temperature in sealed containers with desiccant and protected from light.
Both LS and AzoLS were used within two weeks of preparation.

\subsection*{Encapsulation of \textit{Chlamydomonas} cells into giant liposomes}
An aliquot (2~$\mathrm{\mu L}$) of IS was added to 20~$\mathrm{\mu L}$ of LS in a 1.5-mL microcentrifuge tube.
The tube was vigorously agitated by repeatedly sliding it back and forth across a test-tube rack to form a milky emulsion.
The emulsion was layered over 30~$\mathrm{\mu L}$ of OS in a separate PCR tube.
The PCR tube was centrifuged at 500~$\times g$ for 5~min.
During centrifugation, aqueous droplets containing {\clm} cells passed through the interface between the upper oil phase and the lower aqueous phase (OS), acquired a lipid bilayer, and formed giant liposomes encapsulating {\clm}.
After removal of the upper oil phase, the pelleted {\clmlp}s were collected and resuspended in OS.

\subsection*{Observation of swimming {\clmlp}}
After encapsulation, {\clmlp}s were prepared for observation in a chamber constructed using double-sided tape spacers ($\sim$0.1 mm thick) placed between a glass slide and a coverslip. For osmotic perturbation, an aliquot (8 $\mathrm{\mu L}$) of the {\clmlp} suspension in OS (100 mM glucose) was mixed 1:1 with either a hypertonic glucose solution (180 mM) or a hypotonic glucose solution (20 mM), yielding final glucose concentrations of 140 mM and 60 mM, respectively, before being introduced into the chamber as OS. The chamber was sealed with VALAP to prevent evaporation and observed under a fluorescence microscope (Olympus BX53) equipped with a high-speed camera. Swimming behaviors were recorded at 300--600 fps under red illumination, with the intensity carefully adjusted to minimize photophobic responses in {\clm} cells.
Each data point represents a distinct chlamylipo. In total, 487 chlamylipos were analyzed over seven separate experimental days.

\subsection*{Photoswitching motility control using AzoPC}
The motility of AzoPC-containing {\clmlp}s was switched on and off using brief pulses ($<500$~ms) of UV light (365~nm) and blue light (465~nm).
Green light was used to guide the swimming direction of the {\clmlp}s, UV and blue light were used to switch their motility, and red light was used for observation because {\clm} does not exhibit a phototactic response to red light.
All photoresponsive experiments were conducted using hypotonic OS.
In a separate preliminary test conducted at the same optical power, the cells ceased swimming after approximately 40~s of continuous UV illumination. Because each UV pulse in the photoswitching experiments lasted 500~ms, these exposures were unlikely to abolish algal motility within several switching cycles, although the longer-term effects of repeated UV exposure were not systematically evaluated.

\subsection*{Image processing}
The images were segmented into three classes (background, liposome, and {\clm}) using Otsu's method, as shown in Fig.~S3.
The liposome contour was identified from the fluorescence of calcein contained in the IS.
Samples for which reliable segmentation or tracking was not possible were excluded from quantitative analysis.
Geometric parameters, including the liposome radius $\rl$, the major axis of {\clm} $\rc$, and the protrusion length $p$, were quantified from the segmented images.
The swimming velocity $U$ was calculated from the displacement of the liposome centroid over intervals of $50/600$~s.
The beating frequency $f$ was determined by counting peaks in the temporal variation of the liposome contour.

\renewcommand{\bibsection}{\section*{References}}
\bibliography{reference}

\end{document}